\documentclass{kapproc} 

\usepackage{procps} 

\usepackage[dvips]{graphicx}

\upperandlowercase

\let\footnote\savefootnote

\let\footnotetext\savefootnotetext

\let\footnoterule\savefootnoterule

\kluwerbib 

\begin{document}

\def\ts{\thinspace}                                                       
\def\gapprox{$_>\atop{^\sim}$} \def\lapprox{$_<\atop{^\sim}$}             
\newdimen\sa  \def\sd{\sa=.1em  \ifmmode $\rlap{.}$''$\kern -\sa$         
                                \else \rlap{.}$''$\kern -\sa\fi}          
\newdimen\sb  \def\md{\sb=.06em \ifmmode $\rlap{.}$'$\kern -\sb$         
                                \else \rlap{.}$'$\kern -\sb\fi}          
\def\col#1{\empty}                                                        
\def\bw#1{#1}       

\def\col#1{#1}      
\def\bw#1{\empty}                                                         


\articletitle[Secular Evolution]
{Secular Evolution and the Growth of \\ 
 Pseudobulges in Disk Galaxies}

\author{John Kormendy and Mark E.~Cornell\altaffilmark{1}}
 
\affil{\altaffilmark{1}Department of Astronomy,
                       University of Texas, Austin, TX, USA}

\begin{abstract}

\pretolerance=10000  \tolerance=10000

      Galactic evolution is in transition from an early universe dominated by
hierarchical clustering to a future dominated by secular processes. These result
from interactions involving collective phenomena such as bars, oval disks,
spiral structure, and triaxial dark halos.  A detailed review is in Kormendy \& 
Kennicutt (2004).  This paper provides a summary illustrated in part with
different galaxies.

      Figure 2 summarizes how bars rearrange disk gas into outer rings, inner
rings, and galactic centers, where high gas densities feed starbursts. 
Consistent with this picture, many barred and oval galaxies are observed to 
have dense central concentrations of gas and star formation.  Measurements of
star formation rates show that bulge-like stellar densities are constructed on
timescales of a few billion years.  We conclude that secular evolution builds
dense central components in disk galaxies that look like classical -- that is, 
merger-built -- bulges but that were made slowly out of disk gas.  We call 
these pseudobulges. 

      Many pseudobulges can be recognized because they have characteristics of
disks -- (1) flatter shapes than those of classical bulges,
(2) correspondingly large ratios of ordered to random velocities, (3) small
velocity dispersions $\sigma$ with respect to the Faber-Jackson correlation
between $\sigma$ and bulge luminosity, (4) spiral structure or nuclear bars,
(5) nearly exponential brightness profiles, and (6) starbursts.  These
structures occur preferentially in barred and oval galaxies in which secular 
evolution should be most rapid. Thus a variety of observational and theoretical
results contribute to a new paradigm of secular evolution that complements
hierarchical clustering.

\end{abstract}

\begin{keywords}
Galaxy dynamics, galaxy structure, galaxy evolution
\end{keywords}

\section{Transition From Classical Bulges Built by Hierarchical Clustering
         to Pseudobulges Built by Secular Evolution}

      The relative importance of different processes of galactic evolution
(Fig.~1) is changing as the universe expands.  Rapid processes that happen in
discrete events are giving way to slow, ongoing processes.  Hierarchical
clustering that builds classical bulges is giving way to the secular growth of
pseudobulges.

\begin{figure}[ht!]
\col{\centerline{\includegraphics[width=4.724in] {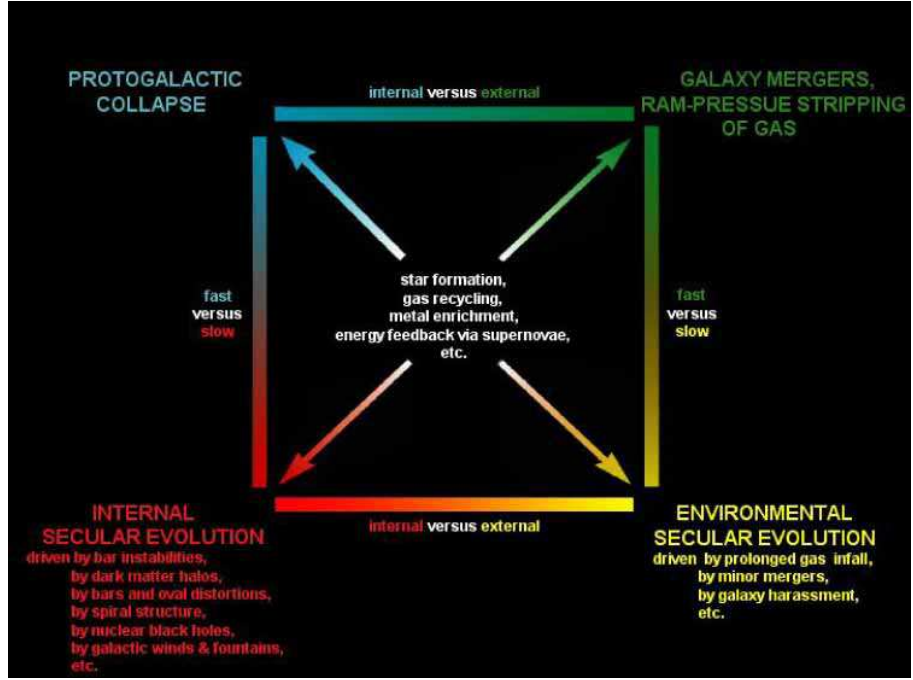}}}
\bw{ \centerline{\includegraphics[width=4.724in] {kormendy_figure1.ps}}}
\caption{Morphological box (Zwicky 1957) of processes of galactic evolution.
Updated from Kormendy (1982a), this figure is from Kormendy \& Kennicutt (2004).
Processes are divided vertically into fast (top) and slow (bottom).  Fast
evolution happens on a free-fall timescale, $t_{\rm ff} \sim (G\ts\rho)^{-1/2}$;
$\rho$ is the density of the object produced and $G$ is the gravitational
constant.  Slow means many galaxy rotation periods.  Processes are divided
horizontally into ones that happen internally in one galaxy (left) and ones 
that are driven by environmental effects such as galaxy interactions (right).
The processes at center are aspects of all types of galaxy evolution.  This
paper reviews the internal and slow processes at lower-left.}
\end{figure}

\pretolerance=10000  \tolerance=10000

      Galactic evolution studies over the past 25 years show convincingly that
hierarchical clustering (see White 1997 and Steinmetz 2001 for reviews) and
mergers (Toomre 1977a, see Schweizer 1990 for a review) built and continue
to build elliptical galaxies and elliptical-like classical bulges of disk
galaxies.  As the universe expands and as galaxy clusters virialize and acquire
large velocity dispersions, mergers get less common (Toomre 1977a; Le Fevre
et al.~2000; Conselice et al.~2003).  Very flat disks in pure disk galaxies
show that at least some galaxies have suffered no major merger violence since
disk star formation began (see Freeman 2000 for a review). Therefore there has 
been time to reshape galaxies via the interactions of individual stars or gas
clouds with collective phenomena such as bars, oval distortions, spiral
structure, and triaxial dark matter halos.  These secular processes are 
reviewed in Kormendy (1993) and in Kormendy \& Kennicutt (2004).  This paper 
provides a summary.

\section{Secular Evolution of Barred Galaxies}

      Why do we think that secular evolution is happening?  The observational
evidence is discussed in \S\ts4, but the story begins with the forty-year
history of simulations of the response of gas to bars.  Figure 2 illustrates
this response and how well it accounts for barred galaxy morphology.  The
angular momentum transfer from bar to disk that makes the bar grow also
rearranges disk gas into outer rings near outer Lindblad resonance (O in the
figure at upper-left), inner rings near bar corotation (C), and dense
concentrations of gas near the center.

\begin{figure}[ht!]
\col{\centerline{\includegraphics[width=4.67in] {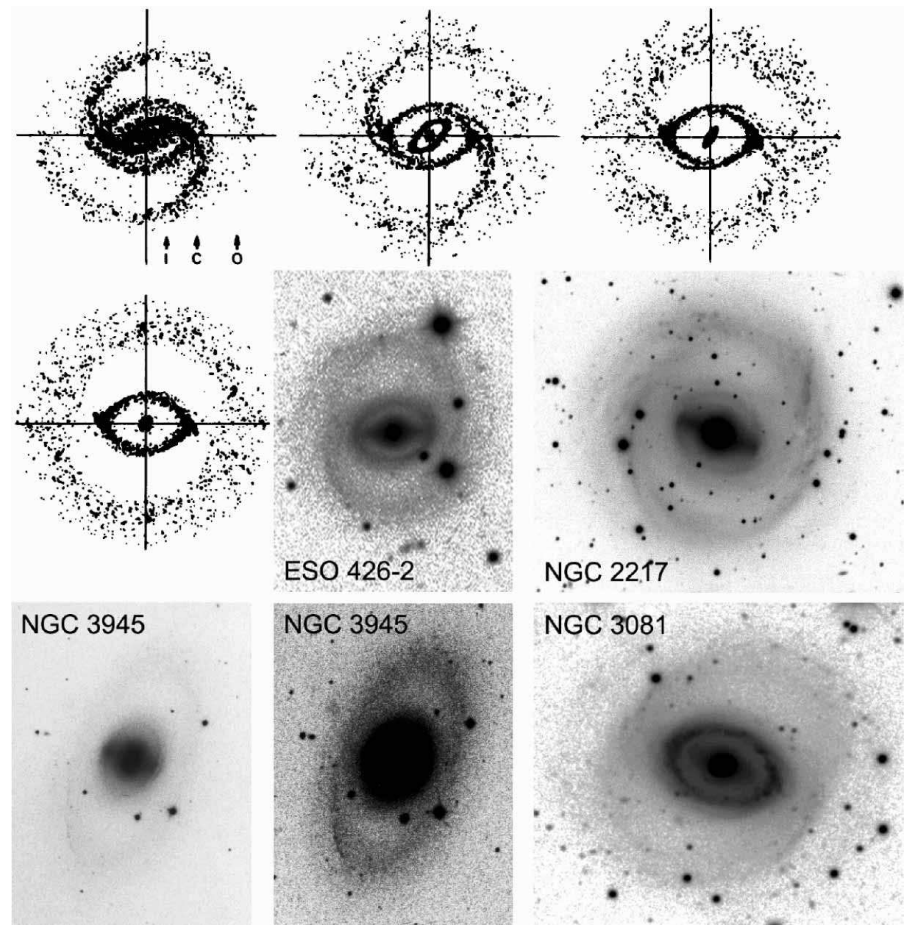}}}
\bw{ \centerline{\includegraphics[width=4.67in] {kormendy_figure2.ps}}}
\caption{Evolution of gas in a rotating oval potential (Simkin, Su, \& Schwarz
1980).  The gas particles in this sticky-particle \hbox{$n$-body} model are
shown after 2, 3, 5, and 7 bar rotations (top-left through center-left).  
Four SB0 or SB0/a galaxies are shown that have outer rings and a lens (NGC 3945)
or an inner ring (most obvious in ESO 426-2 and in NGC 3081).  Sources: 
NGC 3945{\ts}--{\ts}Kormendy (1979); NGC 2217, NGC 3081{\ts}--{\ts}Buta et
al.~(2004); ESO 426-2{\ts}--{\ts}Buta \& Crocker (1991).  This figure is
from Kormendy \& Kennicutt (2004).
}
\end{figure}
\clearpage

   The essential features of Figure 2 are well confirmed by more recent, 
\hbox{state-of-the-art} simulations.
In a particularly important paper, Athanassoula (1992) focuses on the gas 
shocks that are identified with dust lanes in bars.~The shocks are a
consequence of gravitational torques.  Gas accelerates as it approaches and
decelerates as it leaves the potential minimum of the bar.~Therefore it piles
up and shocks near the~ridge line of the bar.  Athanassoula finds that, if the
mass distribution is centrally concentrated enough to result in an inner
Lindblad resonance, then the shocks are offset in the forward (rotation)
direction from the ridge line of the bar.  That is, incoming gas overshoots the
ridge line of the bar before it plows into the departing gas.  The nearly radial
dust lanes seen in bars are essentially always offset in the forward (rotation)
direction.  Compelling support for the identification of the shocks with these
dust lanes is provided by the observation of large velocity jumps across the
dust lanes 
(Pence \& Blackman 1984;
Lindblad, Lindblad, \& Athanassoula 1996;
Regan, Sheth, \& Vogel 1999;
Weiner et al.~2001;
and especially Regan, Vogel, \& Teuben 1997).  

      Shocks inevitably imply that gas flows
toward the center.  Because the shocks are nearly radial, the gas impacts them
almost perpendicularly.  Large amounts of dissipation make the gas sink rapidly.
Athanassoula estimates that azimuthally averaged gas sinking
rates are typically 1 km s$^{-1}$ and in extreme cases up to $\sim 6$ km
s$^{-1}$.  Because 1~km s$^{-1}$ = 1 kpc ($10^9$ yr)$^{-1}$, the implication is
that most gas in the inner part of the disk finds its way to the vicinity of the
center over the course of several billion years, if the bar lives that long.

      Crunching gas likes to make stars.  Expectations from the Schmidt (1959)
law are consistent with observations of enhanced star formation, often in
substantial starbursts near the center.  Examples are shown in Figure 3.  Most
of these are barred galaxies illustrated in Sandage \& Bedke (1994).  NGC 4736
is a prototypical unbarred oval galaxy.  It is included to illustrate
the theme of the next section that barred and oval galaxies evolve similarly.

      Kormendy \& Kennicutt (2004) compile gas density and star formation rate
(SFR) measurements for 20 nuclear star-forming rings.  The SFR densities are
1\ts--\ts3 orders of magnitude higher than the SFR densities averaged over
galactic disks.  Gas densities are correspondingly high:~nuclear star-forming
rings lie on the extrapolation of the Schmidt law, SFR $\propto$ (gas
density)$^{1.4}$.  The BIMA Survey of Nearby Galaxies (SONG) (Regan et al.~2001)
shows that {\it molecular gas densities follow stellar light densities, especially in barred and oval galaxies, even where the stellar densities rise
toward the center above the inward extrapolation of an exponential fitted to
the outer disk.  Since star formation rates rise faster than linearly with
gas density, this guarantees that the observed pseudobulges will grow in
density faster than their associated disks.  That is, pseudobulge-to-disk ratios
increase with time.}  Growth rates to reach the observed stellar densities are
a few billion years.

\begin{figure}[ht!]
\col{\centerline{\includegraphics[width=4.724in] {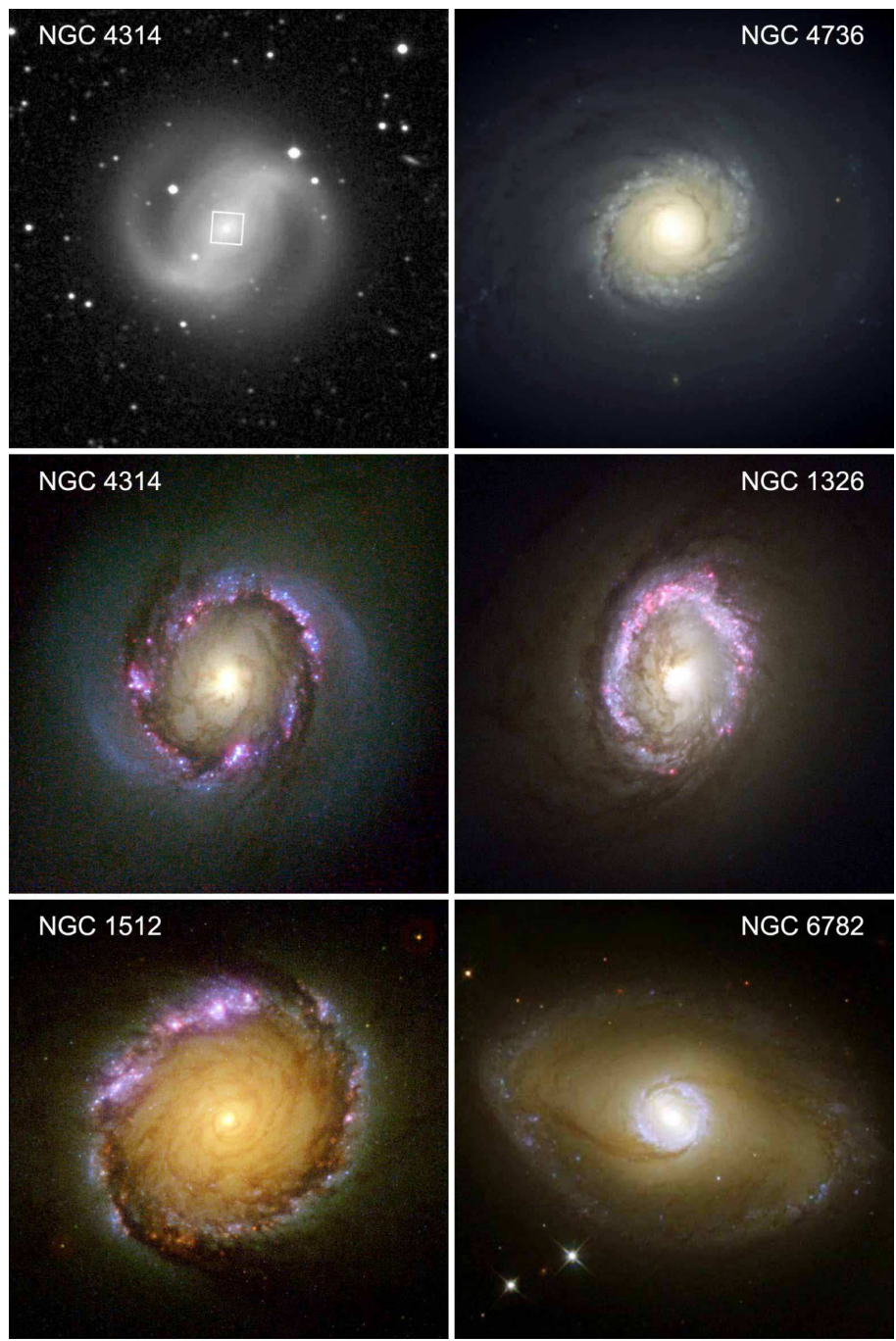}}}
\bw{ \centerline{\includegraphics[width=4.724in] {kormendy_figure3.ps}}}
\caption{Nuclear star formation rings in barred and oval galaxies.  Sources: 
NGC 4314 -- Benedict et al.~(2002); NGC 4736 -- NOAO; NGC 1326 -- Buta et
al.~(2000) and Zolt Levay (STScI); NGC 1512 -- Maoz et al.~(2001); NGC 6782 --
Windhorst et al.~(2002) and the Hubble Heritage Program.  This figure is
from Kormendy \& Kennicutt (2004).
}
\end{figure}
\clearpage

\vfill\eject

\section{Secular Evolution of ``Unbarred'' Galaxies}

      How general are the results of Section 2?  There are four reasons why
we suggest that secular evolution and pseudobulge building are important in 
more than the $\sim 1/3$ of all disk galaxies that look barred at optical
wavelengths:

\begin{enumerate}

\item As emphasized at this conference, near-infrared images penetrate dust
absorption and are insensitive to the low-$M/L$ frosting of young stars in
galaxy disks.  They show us the old stars that trace the mass distribution.  
They reveal that bars are hidden in many galaxies that look unbarred in the
optical
(Block \& Wainscoat 1991;
Spillar et al.~1992;
Mulchaey \& Regan 1997;
Mulchaey et al.~1997; 
Seigar \& James 1998; 
Knapen et al.~2000;
Eskridge et al.~2000, 2002;
Block et al.~2001;
Laurikainen \& Salo 2002;
Whyte et al.~2002).  
About two-thirds of all spiral galaxies look barred in the infrared.
Measures of bar strengths based on infrared images (Buta \& Block 2001; Block 
et al.~2001; Laurikainen \& Salo 2002) should help to tell us the consequences
for secular evolution.

\item Many unbarred galaxies are globally oval.  Ovals are less elongated than
bars{\ts}--{\ts}typical axial ratios are $\sim$\ts0.85 compared with
$\sim$\ts0.2 for bars{\ts}--{\ts}but more of the disk mass participates in the
nonaxisymmetry.  
Strongly oval galaxies can be recognized independently by photometric 
criteria (Kormendy \& Norman 1979; Kormendy 1982a) and by kinematic criteria
(Bosma 1981a, b).  \underbar{Brightness distributions:} The disk consists of 
two nested ovals, each with a shallow surface brightness gradient interior to
a sharp outer edge.  The inner oval is much brighter than the outer one.  The
two ``shelves'' in the brightness distribution have different axial ratios and
position angles, so they must be oval if they are coplanar.  But the flatness 
of edge-on galaxies shows that such disks are oval, not warped.
\underbar{Kinematics:} Velocity fields in oval disks are symmetric and
regular, but (1) the kinematic major axis twists with radius, (2) the optical
and kinematic major axes are different, and (3) the kinematic major and
minor axes are not perpendicular.  Twists in the kinematic principal axes are
also seen when disks warp, but Bosma (1981a, b) points out that warps happen at
larger radii and lower surface brightnesses than ovals.  Also, observations (2) 
and (3) imply ovals, not warps.  Kormendy (1982a) shows that the photometric 
and kinematic criteria for recognizing ovals are in excellent agreement.  

      Oval galaxies are expected to evolve similarly to barred galaxies.  Many
simulations of the response of gas to ``bars'' assumed that all of the potential
is oval rather than that part of the potential is barred and the rest is not.  
NGC 4736 is a prototypical oval with strong evidence for secular evolution
(Figures 3 and 6 here; Kormendy \& Kennicutt 2004).  

\item Bars commit suicide by transporting gas inward and building up the
      central mass concentration 
(Hasan \& Norman 1990; 
Freidli \& Pfenniger 1991;
Friedli \& Benz 1993;
Hasan, Pfenniger, \& Norman 1993;
Norman, Sellwood, \& Hasan 1996;
Heller \& Shlosman 1996;
Berentzen et al.~1998;
Sellwood \& Moore 1999).  
Norman et al.~(1996) grew a point mass at the center of an $n$-body disk that
previously had formed a bar.  As they turned on the point mass, the bar
amplitude weakened.  Central masses of 5\ts--\ts7\ts\% of the disk mass 
dissolved the bar completely.  Shen \& Sellwood (2004) find that central masses
with small radii, like supermassive black holes, destroy bars more easily 
than ones with radii of several hundred parsecs, like pseudobulges.  A bar can
tolerate a soft central mass of 10\ts\% of the disk mass. Observations suggest
that still higher central masses can be tolerated when the bar gets very
nonlinear.  The implication is this: Even if a disk galaxy does not currently
have a bar, bar-driven secular evolution may have happened in the past.

\item  Late-type, unbarred, but global-pattern spirals are expected to evolve
like barred galaxies, only more slowly.  Global spirals are density waves that
propagate through the disk (Toomre 1977b).  In general, stars and gas revolve
around the center faster than the spiral arms, so they catch up to the arms
from behind and pass through them.  As in the bar case, the gas accelerates as
it approaches the arms and decelerates as it leaves them.  Again, the results
are shocks where the gas piles up.  This time the shocks have a spiral shape.
They are identified with the dust lanes on the concave side of the spiral arms
(Figure~4).  Gas dissipates at the shocks, but it does so more weakly than in
barred galaxies, because the gas meets the shock obliquely.  Nevertheless, it
sinks.  In early-type spirals with big classical bulges, the spiral structure
stops at an ILR at a large radius.  The gas may form some stars there, but since
the bulge is already large, the relative contribution of secular evolution is
likely to be minor.  In late-type galaxies, the spiral structure extends close
to the center.  Sinking gas reaches small radii and high densities. We suggest
that star formation then contributes to the building of pseudobulges.
Moreover, late-type galaxies have no classical bulges.  So secular growth of
pseudobulges can most easily contribute a noticeable part of the central mass
concentration precisely in the galaxies where the evolution is most important.

     M{\ts}51 and NGC 4321 (Figure 4) are examples of nuclear star formation in 
unbarred galaxies.  Their exceedingly regular spiral structure and associated
dust lanes wind down close to the center, where both galaxies have bright
regions of star formation (e.{\ts}g., Knapen et al.~1995a, b; Sakamoto et
al.~1995; Garcia-Burillo et al.~1998).  They are examples of secular
evolution in galaxies that do not show prominent bars or ovals.  

     \end{enumerate}

\begin{figure}[ht!]
\col{\centerline{\includegraphics[width=3.7in] {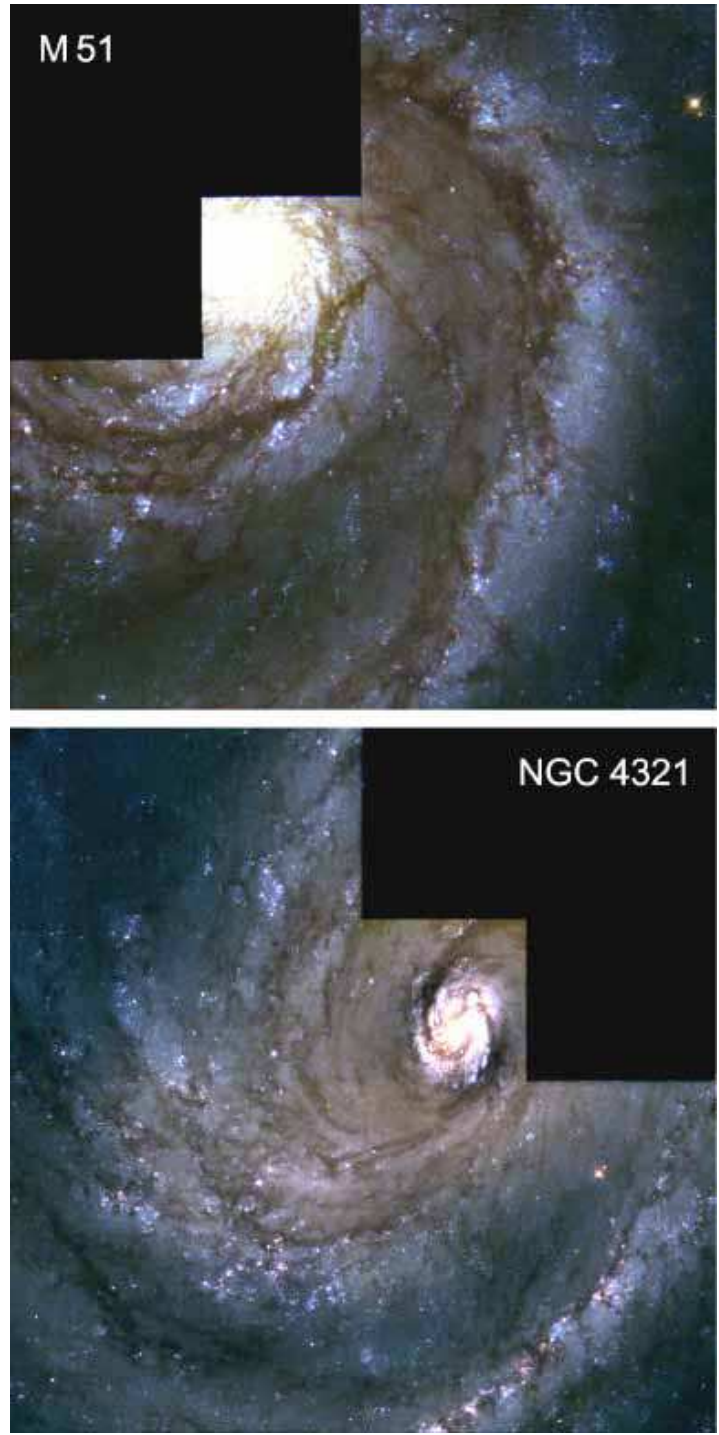}}}
\bw{ \centerline{\includegraphics[width=3.7in] {kormendy_figure4.ps}}}
\caption{Nuclear star formation in the unbarred galaxies M{\ts}51 and NGC
4321 (M{\ts}100).  Dust lanes on the trailing side of the global spiral
arms reach in to small radii.  As in barred spirals, they are are
indicative of gas inflow.  Both galaxies have concentrations of star
formation near their centers that resemble those in Figure 3.  These images
are from the {\itshape Hubble Space Telescope\/} and are reproduced here
courtesy of STScI.}
\end{figure}

\clearpage 

\section{The Observed Properties of Pseudobulges}

     Kormendy (1982a, b) suggested that what we now call pseudobulges were
built by secular inward gas transport and star formation.  Combes \& Sanders (1981) suggested that boxy bulges formed from bars that heat themselves in the
axial direction.  Pfenniger \& Norman (1990) discuss both processes.  These
themes -- dissipational and dissipationless, secular pseudobulge building -- 
have persisted in the literature ever since (see Kormendy \& Kennicutt 2004).

     How can we tell whether a ``bulge'' formed by these processes?
Fortunately, pseudobulges retain enough memory of their disky origin so that 
the best examples are recognizable.  Structural features that indicate a disky
origin include nuclear bars, nuclear disks, nuclear spiral structure, boxy
bulges, exponential bulges, and central star formation  (Figures 3 and 4).
We consider all of these to be features of pseudobulges, because the evidence
is that all of them are built secularly out of disk material.  Similarly,
global spiral structure, flocculent spiral structure, and no
spiral structure in S0 galaxies are all features of disks.  In addition,
pseudobulges are more dominated by rotation and less dominated by random motions
than are classical bulges and ellipticals.  

      Spectacular progress in recent years has come from {\itshape HST\/}
imaging surveys.   We begin with these surveys.  To make data available on more
galaxies, we also provide a detailed discussion of two galaxies, 
NGC 4371 and NGC 3945, that are different from the ones discussed in Kormendy
\& Kennicutt (2004).

\subsection{Embedded Disks: Spiral Structure, Star Formation}

     Renzini (1999) states the definition of a bulge: ``It appears legitimate 
to look at bulges as ellipticals that happen to have a prominent disk around
them [and] ellipticals as bulges that for some reason have missed the
opportunity to acquire or maintain a prominent disk.''  Our paradigm of galaxy
formation is that bulges and ellipticals both formed via galaxy mergers 
(e.{\ts}g., Toomre 1977a; Steinmetz \& Navarro 2002, 2003), a
picture that is well supported by observations (see Schweizer 1990 for a
review).  But as observations improve, we discover more and more features that
make it difficult to interpret every example of what we used to call a ``bulge''
as an elliptical living in the middle of a disk. Carollo and collaborators find
many such galaxies in their {\itshape HST\/} snapshot survey of 75,
S0{\ts}--{\ts}Sc galaxies observed with WFPC2 in $V$ band (Carollo et al.~1997,
1998; Carollo \& Stiavelli 1998; Carollo 1999) and a complementary survey of 78
galaxies observed with NICMOS in $H$ band (Carollo et al.~2001, 2002; Seigar et
al.~2002).  Figure 5 shows examples.  These are Sa{\ts}--{\ts}Sbc galaxies,
so they should contain bulges.  Instead, their centers look like star-forming
spiral galaxies.  It is difficult to believe that, based on such images, anyone
would define bulges as ellipticals living in the middle of a disk.  Spiral
structure happens only in a disk. Therefore these are examples of pseudobulges.

\begin{figure}[ht!]
\col{\centerline{\includegraphics[width=4.724in] {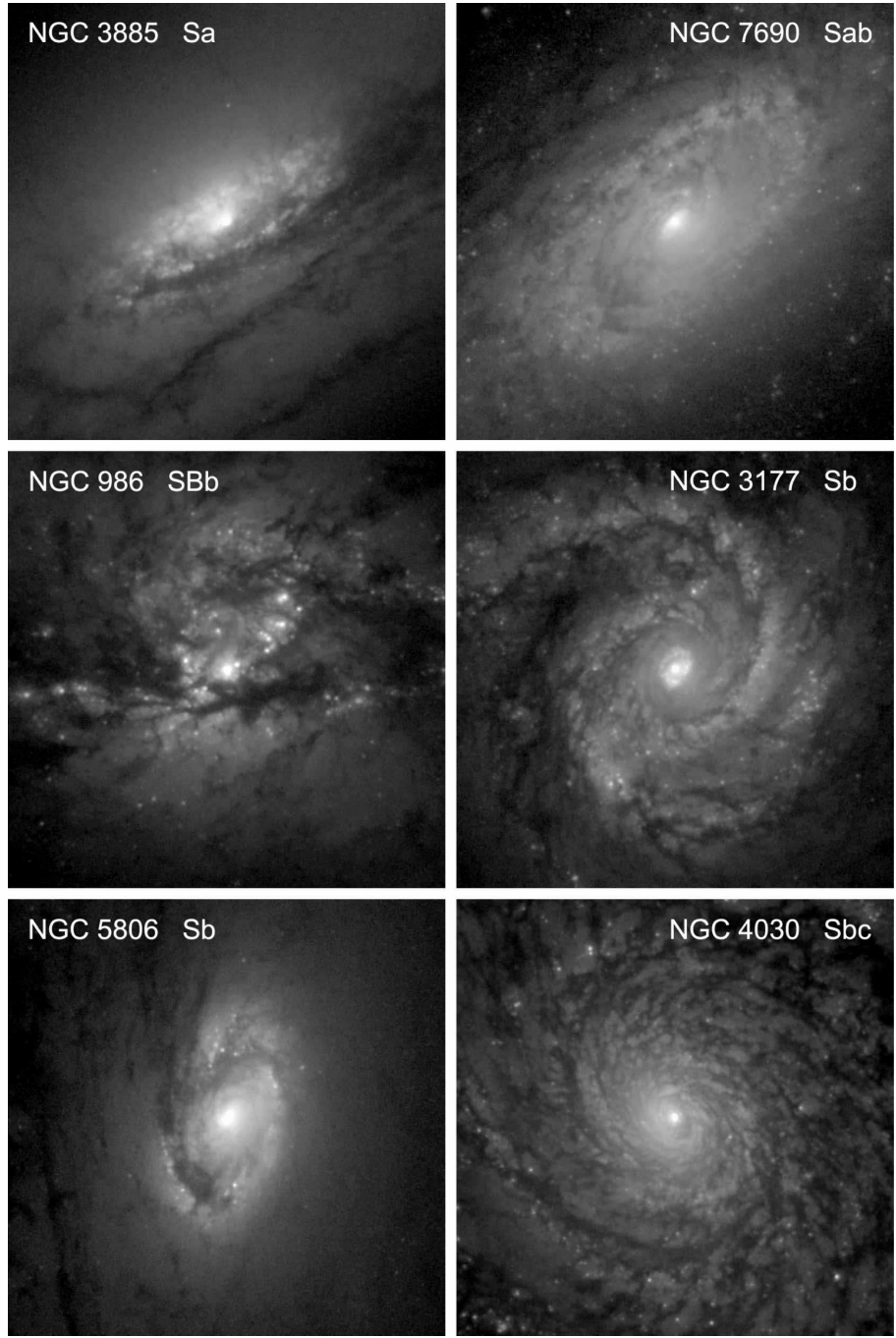}}}
\bw{ \centerline{\includegraphics[width=4.724in] {kormendy_figure5.ps}}}
\caption{Sa -- Sbc galaxies whose ``bulges'' have disk-like morphology.  Each
panel shows an 18$^{\prime\prime}$ $\times$ 18$^{\prime\prime}$ region centered
on the galaxy nucleus and extracted from {\itshape HST\/} WFPC2 F606W images
taken and kindly provided by Carollo et al.~(1998).  North is up and east is 
at left.  Displayed intensity is proportional to the logarithm of the galaxy
surface brightness.}
\end{figure}

\clearpage

\subsection{Rotation-Dominated Pseudobulges}

    Figure 6, the $V_{\rm max}/\sigma${\ts}--{\ts}$\epsilon$ diagram
(Illingworth 1977; Binney 1978a, b), shows that pseudobulges (filled
symbols) are more rotation-dominated than classical bulges (open symbols)
which are more rotation-dominated than giant ellipticals (crosses).  This is
disky behavior, as follows.  Seen edge-on, rotation-dominated disks have
parameters that approximately satisfy the extrapolation of the oblate line to
$\epsilon \geq 0.8$.  Observed other than edge-on, they project well above the
oblate line.  In contrast, projection keeps $\epsilon$ \lapprox \ts\ts0.6 
isotropic spheroids near the oblate line.  The filled symbols therefore
represent objects that contain rapidly rotating and hence disky central
components.  Of the most extreme cases, NGC 4736 is discussed in detail in
Kormendy \& Kennicutt (2004).  Complementary photometric evidence for
pseudobulges in NGC 3945 and NGC 4371 is discussed in the next subsection.

\begin{figure}[ht!]
\col{\centerline{\includegraphics[width=3.4in] {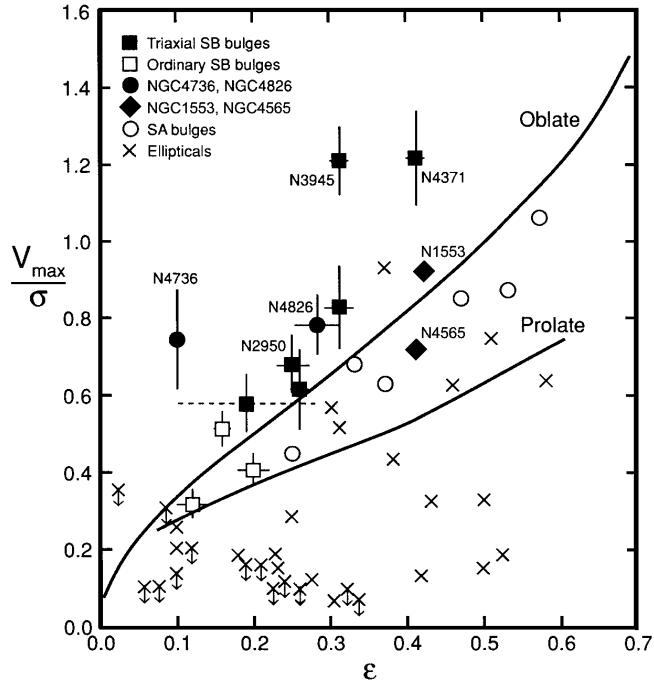}}}
\bw{ \centerline{\includegraphics[width=3.4in] {kormendy_figure6.ps}}}
\caption{Relative importance of rotation and velocity dispersion: 
$V_{\rm max}/\sigma$ is the ratio of the maximum rotation
velocity to the mean velocity dispersion interior to the half-light radius;
$(V_{\rm max}/\sigma)^2$ measures the relative contribution of ordered and
random motions to the total kinetic energy and hence, via the virial theorem,
to the dynamical support that gives the system its ellipticity $\epsilon$
(Binney \& Tremaine 1987).  The ``oblate'' line describes oblate spheroids
that have isotropic velocity dispersions and that are flattened only by
rotation.  The ``prolate'' line is one example of how prolate spheroids can
rotate more slowly for a given $\epsilon$ because they are flattened partly by
velocity dispersion anisotropy.  This figure is from Kormendy\& Kennicutt
(2004).}
\end{figure}

\clearpage

\subsection{Embedded Disks -- II.~Flat Pseudobulges}

      That some pseudobulges are essentially as flat as disks is inferred when
we observe spiral structure (Figure 5), but it is observed directly in
surface photometry of highly inclined galaxies (see Kormendy 1993 and Kormendy
\& Kennicutt 2004 for reviews).  Figures 7{\ts}--{\ts}9 show examples.

  The SB0 galaxy NGC 4371 contains one of the most rotation-dominated 
``bulges'' in Figure 6.  This result (Kormendy 1982b, 1993) already implies 
that NGC 4371 contains a pseudobulge.

\begin{figure}[ht!]
\col{\centerline{\includegraphics[width=4.69in]  {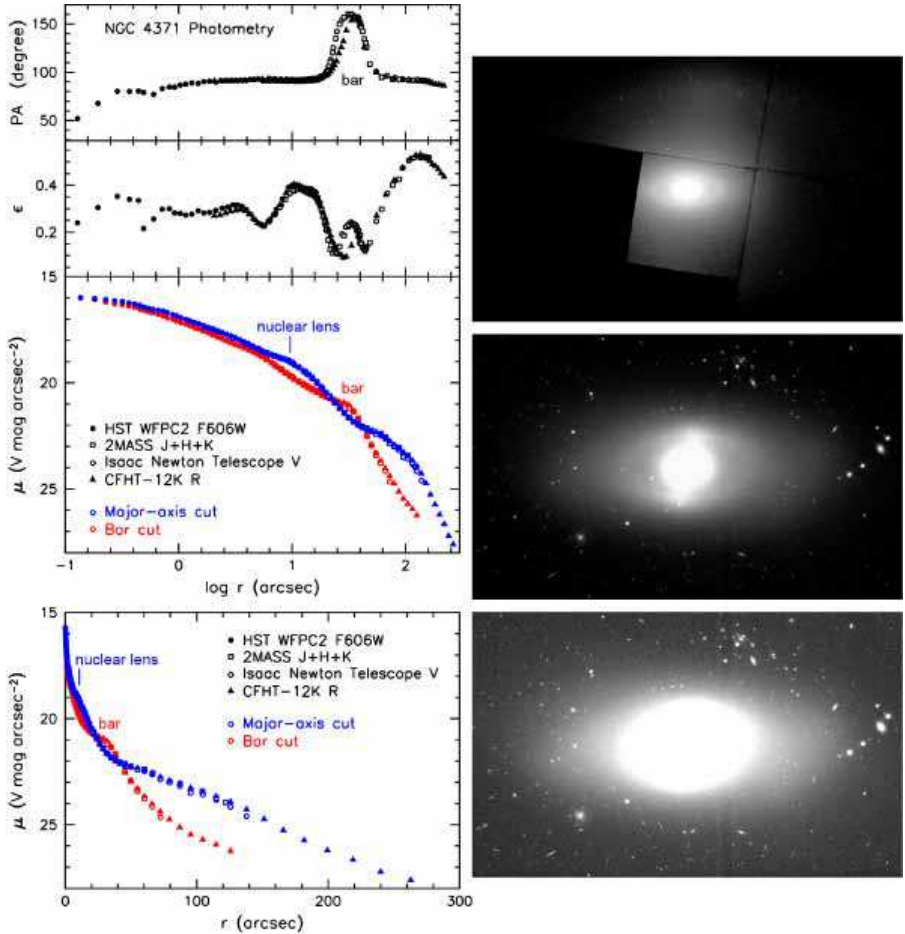}}}
\bw{ \centerline{\includegraphics[width=4.724in] {kormendy_figure7.ps}}}
\caption{NGC 4371 pseudobulge -- top image: 130$^{\prime\prime}$ 
$\times$ 80$^{\prime\prime}$ WFPC2 F606W image from the {\itshape HST\/}
archive; middle and bottom: 406$^{\prime\prime}$ $\times$
249$^{\prime\prime}$ CFHT 12K $R$-band images from Kormendy et al.~(2004)
at different logarithmic intensity stretches.  North is up and east is at left.
The plots show surface photometry, including brightness cuts along the major 
and bar axes (see the text) shifted to the $V$-band INT zeropoint derived using
aperture photometry from Poulain (1988).}
\end{figure}

\clearpage

\begin{figure}[ht!]   
\col{\centerline{\includegraphics[width=3.0in] {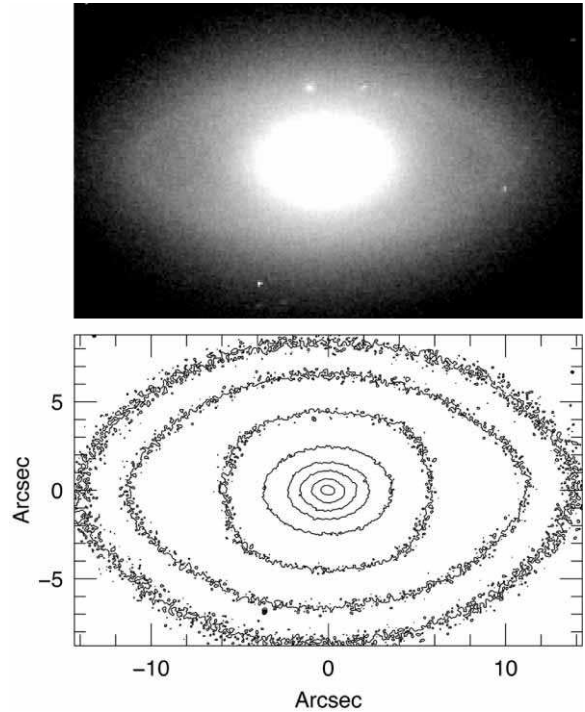}}}
\bw{ \centerline{\includegraphics[width=3.0in] {kormendy_figure8.ps}}}
\caption{NGC 4371 nuclear lens. North is up and east is at left.  The 
{\itshape HST\/} PC F606W image at the top is 28\sd7 $\times$ 17\sd6.  
The intensity stretch is logarithmic.  The bottom panel shows isophotes chosen
to distinguish the high-ellipticity lens from the rounder center. The contour
levels are 19.6, 19.1, 18.6, 18.0, 17.6, 17.3, 16.8, and 16.3 $V$ mag arcsec$^{-2}$.    Compare the lens in NGC 1553 (Freeman 1975; Kormendy 1984).}
\end{figure}

      Photometry strengthens the evidence that the ``bulge'' is disky. Kormendy
(1979) concluded that ``the spheroid is distorted into a secondary bar;
i.{\ts}e., is prolate''.  Wozniak et al.~(1995) saw this, too, but noted that
it could be a projection effect due to high inclination. Based on unsharp-masked
images, Erwin \& Sparke (1999) conclude that NGC 4371 contains a smooth nuclear
ring and identify this -- in effect, if not in name -- as a pseudobulge. 

      Figure 7 shows our photometry. The ellipticities $\epsilon$ and position
angles PA are based on ellipse fits to the isophotes. However, the isophotes 
are far from elliptical at some radii, so the bottom two panels show brightness
cuts in 25$^\circ$ wedges along the major and bar axes.  The bar is obvious as 
a shelf in surface brightness and as corresponding features in the $\epsilon$
and PA profiles. Interior to the outer exponential disk shown by the major-axis
cut is a steep central rise in surface brightness that would conventionally be
identified as the bulge.  However, its properties are distinctly not bulge-like.
It contains a shelf in surface brightness with radius $r \simeq
10^{\prime\prime}$; this is shown in more detail~in~Figure~8.  The outer
rim of a shelf looks like a ring when an image is divided by a smoothed version
of itself.  The shelf has the brightness profile of a lens (cf.~the prototype 
in NGC 1553: Freeman 1975; Kormendy 1984).  We interpret it as a nuclear lens. 
The important point is this: The nuclear lens has essentially the same apparent
flattening and position angle as the outermost disk.  We cannot tell from Figure
7 whether it really is a disk or whether it is thicker than a disk and therefore
prolate (a nuclear bar).  Rapid~rotation~(Figure~6) makes the disk
interpretation more likely.  In either case, the nuclear lens is diagnostic of
a pseudobulge (see Kormendy \& Kennicutt 2004 for further discussion).
Photometric criteria (Figures 7 and 8) and dynamical criteria (Figure 6) for 
identifying pseudobulges agree very well in NGC 4371.

      The same is true in NGC 3945 (Figure 9).  As in NGC 4371, the bar of this
SB0 galaxy is oriented almost along the apparent minor axis.  Therefore, when
the ``bulge'' at $r \simeq 10^{\prime\prime}$ looks essentially as flat as the 
outermost disk, there is ambiguity about whether the inner structure is flat
and circular or axially thick and a nuclear bar.  It was interpreted as a
nuclear bar in Kormendy (1979) and in Wozniak et al.~(1995) and is illustrated
as such in Kormendy \& Kennicutt (2004).  In contrast, Erwin \& Sparke (1999)
interpret it as ``probably intrinsically round and flat -- an inner disk''. 
Erwin et al.~(2003) reach the same conclusion in a detailed photometric study.
Based on $\epsilon$ and PA profiles, they also identify an ``inner bar'' with 
radius 2$^{\prime\prime}$.  All of these features are well confirmed by our
photometry (Figure 9).  The main and ``inner'' bars are clear in the $\epsilon$
and PA profiles. The main bar also makes an obvious shelf in the bar-axis
brightness cut, while the nuclear bar is so subtle that we regard it as an
interpretation rather than a certainty.  Our photometry is consistent
with the interpretation either that the shelf at $r \simeq 10^{\prime\prime}$ 
in the major-axis cut is a nuclear bar (in which case the galaxy has
three nested bars) or that this is a nuclear lens which is nearly circular and
very flat.  For the purposes of this paper, we do not have to decide between
these alternatives.  Either one is characteristic of a disk.  Consistent with
the conclusions of all of the above papers, either interpretation implies that
the central rise in surface brightness above the galaxy's primary lens and outer
ring is caused by a pseudobulge.  This may have been added to a pre-existing
classical bulge, but if so, the classical bulge dominates the light only in the
central 1\sd5 (Figure~9; Erwin et al.~2003).

      Present-day gas inflow and star-formation rates imply that
dissipative secular evolution should be most important in Sbc{\ts}--{\ts}Sc
galaxies (Kormendy \& Kennicutt 2004).  Classical bulges are the rule in Sas,
but it is remarkable how easily one can find S0s with pseudobulges.  
We interpret this result as additional evidence for van den Bergh`s (1976)
``parallel sequence'' classification, which recognizes that some S0s have
smaller (pseudo)bulge-to-disk luminosity ratios than do Sa galaxies.  The hint
is that the secular evolution happened long ago, when the galaxies were gas-rich
and before they were converted to S0s.

\begin{figure}[ht!]
\col{\centerline{\includegraphics[width=4.724in] {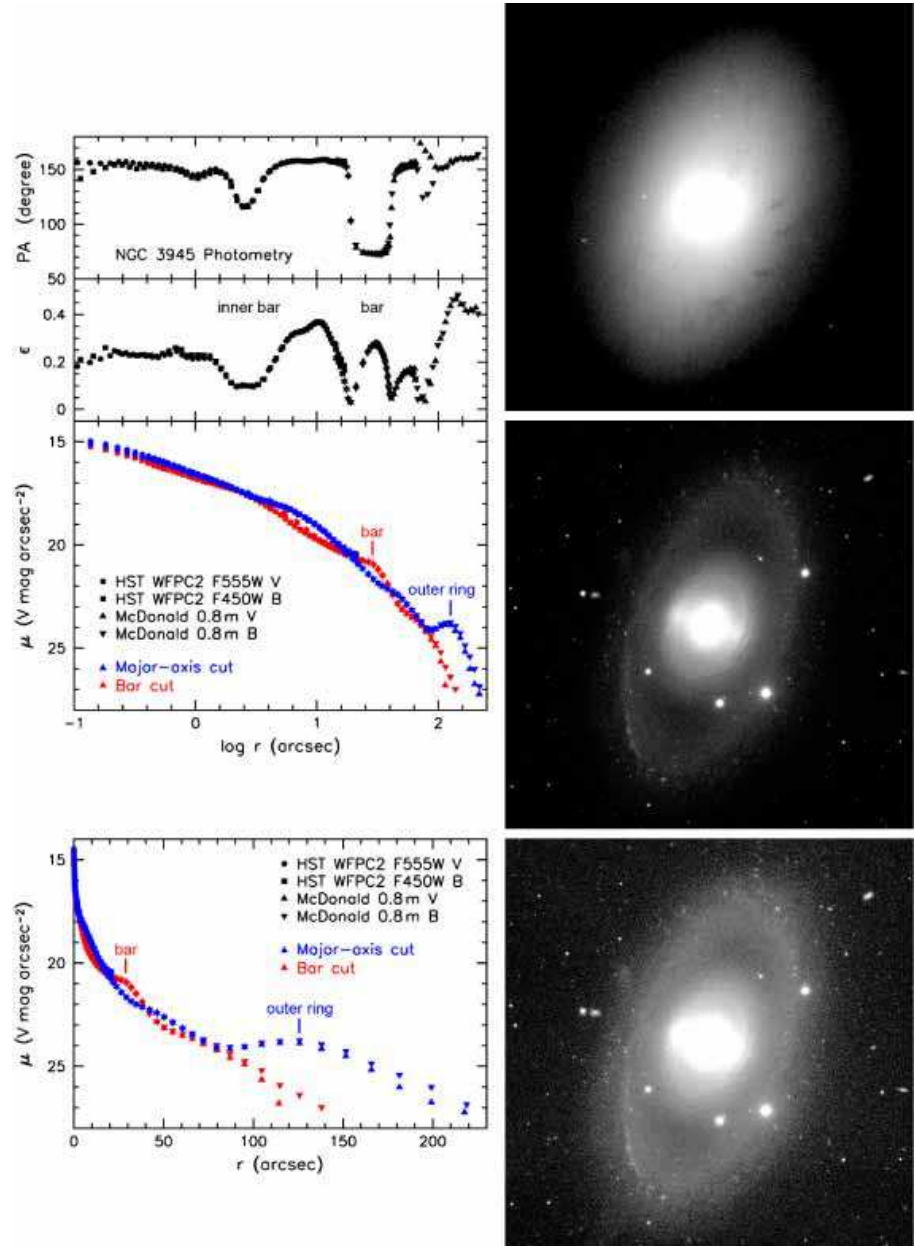}}}
\bw{ \centerline{\includegraphics[width=4.724in] {kormendy_figure9.ps}}}
\caption{NGC 3945 pseudobulge -- top image: 29$^{\prime\prime}$ 
$\times$ 29$^{\prime\prime}$ PC F450W image from the {\itshape HST\/} archive;
middle and bottom: 6\md1 $\times$ 6\md1 WIYN $B$-band images from Buta et
al.~(2004) at different logarithmic intensity stretches.  North is up and east
is at left.  The plots show ellipticity and position angle profiles from 
ellipse fits to the isophotes and major- and bar-axis cuts in 28$^{\circ}$
wedges.  The McDonald Observatory 0.8 m telescope $V$-band images were 
zeropointed using aperture photometry from Burstein et al.~(1987) and from
Angione (1988).  The other profiles are shifted to this zeropoint.  Outer rings
are usually elongated perpendicular to bars, so the apparent ellipticity of
flat, circular isophotes is likely to be the one observed at the largest radii.}
\end{figure}

\clearpage 

\pretolerance=10000  \tolerance=10000

\section{Preliminary Prescription for Recognizing Pseudobulges}

      We have space in this paper to review only a few features of pseudobulges.
Kormendy (1993) and Kormendy \& Kennicutt (2004) discuss others.  In this
section, we list these other features to provide a preliminary prescription 
for identifying pseudobulges.  

      Any prescription must recognize that we expect a continuum from
classical, merger-built bulges through objects with some E-like and some
disk-like characteristics to pseudobulges built completely by secular processes.
Uncertainties are inevitable when we deal with transition objects. Keeping these
in mind, a list of pseudobulge characteristics includes: 

\begin{enumerate}

\item The candidate pseudobulge is seen to be a disk in images:~it shows spiral
structure or its apparent flattening is similar to that of the outer disk. 

\item It is or it contains a nuclear bar (in face-on galaxies).  Bars are disk
      phenomena; they are fundamentally different from triaxial ellipticals.

\item It is box-shaped (in edge-on galaxies).  Box-shaped bulges are intimately
      related to bars; they are believed to be -- or to be made by -- edge-on
      bars that heated themselves in the axial direction.

\item It has $n \simeq 1$ to 2 in a Sersic (1968) function, $I(r) 
      \propto e^{-K[(r/r_e)^{1/n} - 1]}$, fit to the brightness profile.
      Here $n = 1$ for an exponential, $n = 4$ for an
      $r^{1/4}$ law, and $K(n)$ is chosen so that radius $r_e$ contains half 
      of the light in the Sersic component.  Nearly exponential profiles
      prove to be a characteristic of many pseudobulges (e.{\ts}g., 
      Andredakis \& Sanders 1994; Andredakis, Peletier, \& Balcells 1995;
      Courteau, de Jong, \& Broeils 1996; Carollo et al.~2002; 
      Balcells et al.~2003; MacArthur, Courteau, \& Holtzman 2003;
      Kormendy \& Kennicutt 2004).

\item It is more rotation-dominated than are classical bulges in the
      \hbox{$V_{\rm max}/\sigma$ -- $\epsilon$} diagram; e.{\ts}g., 
      $V_{\rm max}/\sigma$ is larger than the value on the oblate line.

\item It is a low-$\sigma$ outlier in the Faber-Jackson (1976) correlation
      between (pseudo)bulge luminosity and velocity dispersion.  

\item It is dominated by Population I material (young stars, gas, and
dust), but there is no sign of a merger in progress. 

\end{enumerate}

      If any of these characteristics are extreme or very well developed,
it seems safe to identify the central component as a pseudobulge.  The more
of 1 -- 7 apply, the safer the classification becomes.  

      Small bulge-to-total luminosity ratios $B/T$ do not guarantee that a
galaxy contains a pseudobulge, but if $B/T$ \gapprox \ts1/3 to 1/2,
it seems safe to conclude that the galaxy contains a classical bulge.

      Based on these criteria, galaxies with classical bulges include M{\ts}31, 
M{\ts}81, NGC 2841, NGC 3115, and NGC 4594.  Galaxies with prototypical pseudobulges include 
NGC 3885 (Figure 5),
NGC 3945 (Figures 2, 6, 9),
NGC 4314 (Figure 3),
NGC 4321 (Figure 4),
NGC 4371 (Figures 6, 7, 8), and
NGC 4736 (Figures 3, 6).
The classification of the bulge of our Galaxy is ambiguous; the observation that
it is box shaped strongly favors a pseudobulge, but stellar population data
are most easily understood if the bulge is classical.

\section{Perspective}

      Internal secular evolution complements environmental processes such as
hierarchical clustering and harrassment in shaping galaxies.  Thirty
years ago, Hubble classification was in active use, but we also knew of
a long list of commonly observed features in disk galaxies, including
lens components, boxy bulges, nuclear bars, and central starbursts, and
also a list of unique peculiar galaxies (e.g., Arp 1966) that were
unexplained and not included in morphological classification schemes.
Almost all these features and peculiar galaxies now have candidate
explanations within one of two paradigms of galaxy evolution that
originated in the late 1970s.  The peculiar objects have turned out
mostly to be interacting and merging galaxies.  And many previously
unexplained features of disk galaxies are fundamental to our
understanding that galaxies evolve secularly long after the spectacular
fireworks of galaxy mergers, starbursts, and their attendant nuclear
activity have died down.

\begin{acknowledgments}

     It is a pleasure to thank Ron Buta and Marcella Carollo for providing many
of the images used in the figures.  David Fisher kindly took the images of NGC
3945 with the McDonald Observatory 0.8 m telescope, and Tom Jarrett provided 
the star-removed 2MASS images of NGC 3945.  This paper is based partly on
observations made with the NASA/ESA {\itshape Hubble Space Telescope}, obtained
from the data archive at the Space Telescope Science Institute.  STScI is
operated by AURA, Inc. under NASA contract NAS 5-26555.  We also used the
NASA/IPAC Extragalactic Database (NED), which is operated by JPL and Caltech
under contract with NASA.

\end{acknowledgments}

\end{document}